\documentclass[11pt,a4paper,twoside]{article}
\usepackage{amsmath}
\usepackage{amsfonts}
\usepackage{changes}
\usepackage{color}
\usepackage{hyperref}
\numberwithin{equation}{section}

\begin{document}

\noindent

{\bf
{\Large Holographic reduction of Maxwell-Chern-Simons \\theory\\ 

}
} 

\vspace{.5cm}
\hrule

\vspace{1cm}

\noindent

{\large\bf{Nicola Maggiore\footnote{\tt nicola.maggiore@ge.infn.it }
\\[1cm]}}

\setcounter{footnote}{0}

\noindent
{{}Dipartimento di Fisica, Universit\`a di Genova,\\
via Dodecaneso 33, I-16146, Genova, Italy\\
and\\
{} I.N.F.N. - Sezione di Genova\\
\vspace{1cm}

\noindent
{\tt Abstract~:}
The 3D Maxwell-Chern-Simons theory with planar, single-edged, boundary is considered. It is shown that its holographic reduction on a flat euclidean 2D spacetime is a scalar field theory describing a conserved chiral current, which corresponds to the electric continuity equation. Differently from the theory with planar, double-edged boundary, in this holographic bulk-boundary approach physical quantities like the chiral velocity of the edge excitations depend also on the non-topological Maxwell term.

\newpage

\section{Introduction}

The introduction of a boundary in a quantum field theory (QFT) is known since long time to yield fruitful consequences. 
In \cite{Symanzik:1981wd}, Symanzik proposed a way to introduce a boundary in a renormalizable quantum field theory with the aim of studying the Casimir effect. In his pioneering work, the presence of boundary is formally translated into a somehow obvious request on the propagators: in a QFT with boundary the propagators must ``separate'', $i.e.$ propagators connecting two points lying on different sides of the boundary must vanish. The presence of the boundary is reduced to a condition on the two-points Green functions, which can be solved only if certain boundary condition are verified. This also solves the puzzle of the ambiguity related to the boundary conditions on the fields, which are not given by hand, but are uniquely determined by the separability conditions. Symanzik's method has been applied, in particular, to investigate what happens on the boundary of topological quantum field theories (TQFTs),  like the three dimensional Chern-Simons (CS) theory \cite{Blasi:2008gt}, and the BF theories \cite{Amoretti:2012kb,Amoretti:2013nv,Amoretti:2014iza}, which live in any spacetime dimensions. The presence of a boundary in a TQFT is particularly relevant. As it is well known, the observables of TQFTs are globally and not locally defined, in the sense that they do not depend on the metric of the manifold on which the theory is defined \cite{Birmingham:1991ty}. The only way to give  local degrees of freedom to a TQFT is to put a boundary on the manifold and see what happens on it. One of the first, and most important, examples of this way of exploiting the  boundary in a TQFT is the classification of all rational conformal field theories by means of a CS theory built on a manifold with boundary \cite{Moore:1989yh}. Moreover, Symanzik's general approach to QFT with boundary has been successfully applied to deduce the conserved chiral current living on the edge in the states of the Fractional Quantum Hall Effect \cite{Blasi:2008gt}. Another important result of Symanzik's method is the description of the experimentally observed  topological insulators \cite{Cho:2010rk}, 
characterized by the presence of conserved currents with opposite chirality and spin on the two edges of the boundary, by means of the topological BF theory, which respects time reversal invariance \cite{Blasi:2011pf}. 
The method of separability of propagators has been applied also to the non-topological Maxwell-Chern-Simons (MCS) theory with planar boundary \cite{Blasi:2010gw}. The outcome has been that the boundary resulted completely transparent to the non-topological Maxwell term: in other words, CS and MCS theory with planar, double-edged, boundary are equivalent. 

Now, the question might arise whether, and how, the above results depend on the type of boundary which is introduced in the theory. Symanzik's separability condition assumes the existence of a double-edged boundary, with left and right hand sides. Another, alternative, example in theoretical physics where the boundary plays a primary role is the AdS/CFT correspondence \cite{Maldacena:1997re,Witten:1998zw,Maldacena:2003nj}, a.k.a. gauge/gravity duality. Initially introduced in the framework of string theory, in more recent times this duality found a second youth in condensed matter physics \cite{Hartnoll:2009sz,McGreevy:2009xe,Zaanen:2015oix,Ammon:2015wua,Amoretti:2017xto}. The gauge/gravity duality concerns a D+1-dimensional bulk, where an Einstein-Hilbert action on an anti-de Sitter (AdS) background is built, and its D-dimensional Minkowskian boundary, where a Conformal Field Theory (CFT) realizes its holographic reduction. The outcome can be encoded in a gauge/gravity dictionary \cite{Natsuume:2014sfa}, which relates D+1-dimensional quantities living in the bulk to their D-dimensional boundary counterparts. The gauge/gravity theories are field theories with boundary which cannot be described {\it \`a la} Symanzik by separating a right hand side from a left hand side. Rather, the boundary is single-edged, and an alternative approach must be adopted. The aim of this paper is to study the holographic limit of MCS theory, realizing the holography in flat spacetime described in \cite{McGreevy:lectures, Amoretti:2014kba}.

The paper is organized as follows. 
In Section 2 the MCS action with boundary is introduced, together with the gauge-fixing and the external sources term, following the standard procedure of gauge field theory. In addition, the most general action living on the boundary is built.
In Section 3 the equations of motion of the MCS theory with boundary are computed, from which the boundary conditions and the Ward identity are derived. From this latter, going on-shell, the existence of a conserved current is deduced, and its physical interpretation is discussed.
In Section 4 the holographic link to the 2D scalar field theory is performed, first by deriving the 2D algebra satisfied by the gauge fields  restricted on the boundary, then writing it in terms of the scalar fields induced by the continuity equation described in the previous Section. The algebra can then be interpreted as commutation relation between canonically conjugate variables, which, in turn, lead to a 2D scalar field theory, holographically dual to the bulk 3D MCS theory. The compatibility between the dual theories is realized by imposing that the 2D equation of motion for the boundary scalar field coincides with the 3D boundary conditions of the bulk MCS gauge fields.
The results of this paper are summarized in the concluding Section 5.

\section{The action with boundary}

The Maxwell-Chern-Simons (MCS) action on the flat three dimensional euclidean spacetime, described by the metric  $\eta_{\mu\nu}=\mbox{diag}(1,1,1)$ is given by\footnote{Our conventions concerning greek and latin indices are as follows: 
$\mu=0,1,2\ ;\ i=0,1$. The Levi-Civita completely antisymmetric tensor is $\epsilon_{\mu\nu\rho}$, with $\epsilon_{012}=1$ and $\epsilon_{2ij}\equiv\epsilon_{ij}$.}
\begin{equation}
S_{MCS}=\alpha S_{CS}+\beta S_{M}\ ,
\label{2.1}\end{equation}
where the Chern-Simons (CS) and the Maxwell terms are, respectively,
\begin{equation}
S_{CS}=\int d^3x\ \epsilon_{\mu\nu\rho}A_\mu\partial_\nu A_\rho 
\label{2.2}\end{equation}
and
\begin{equation}
S_{M}=\int d^3x\ F_{\mu\nu} F_{\mu\nu} \ .
\label{2.3}\end{equation}
The electromagnetic field strength is, as usual, defined by 
$F_{\mu\nu}=\partial_\mu A_\nu - \partial_\nu A_\mu$, and the two constants $\alpha$ and $\beta$ will help us to identify the contributions of the CS and Maxwell terms, respectively.  The canonical mass dimensions of fields and parameters are: 
\begin{equation}
[A]=1/2,\ ;\  [\alpha]=1\ ;\ [\beta]=0\ .
\label{2.4}\end{equation}
As it is well known \cite{Deser:1981wh}, in 3D the CS coupling $\alpha$ plays the role of topological mass for the gauge field $A_\mu(x)$.

The theory is given a planar boundary at $x_2=0$, introduced by means of  the Heaviside step function 
$\theta(x_2)$, so that the  bulk MCS action with boundary is:
\begin{eqnarray}
S_{bulk}&=&
\int d^3x\ \theta(x_2)\left(
\alpha\epsilon_{\mu\nu\rho}A_\mu\partial_\nu A_\rho 
+ \beta F_{\mu\nu} F_{\mu\nu} \right)
\nonumber\\
&=& 
\alpha\int d^3x\ \theta(x_2)
\left[
A_0(\partial_1A_2-\partial_2A_1) - A_1(\partial_0A_2-\partial_2A_0) \right.
\nonumber \\
&&\left. +A_2(\partial_0A_1-\partial_1A_0)
\right]
\label{2.5}\\
&&+
2\beta\int d^3x\ \theta(x_2)
\left[
\partial_0A_1\partial_0A_1
-2 \partial_0A_1\partial_1A_0
+ \partial_1A_0\partial_1A_0
+ \partial_0A_2\partial_0A_2
\right.\nonumber\\
&&\left.
-2 \partial_0A_2\partial_2A_0
+ \partial_2A_0\partial_2A_0
+ \partial_1A_2\partial_1A_2
- 2 \partial_1A_2\partial_2A_1
+ \partial_2A_1\partial_2A_1 \nonumber
\right]
\end{eqnarray}
The action $S_{MCS}$ \eqref{2.1} is left invariant under the infinitesimal gauge transformation 
\begin{equation}
\delta_{gauge} A_\mu=\partial_\mu\lambda(x),
\label{2.6}\end{equation}
where $\lambda(x)$ is a local gauge parameter:
\begin{equation}\delta_{gauge} S_{MCS}=0\ .
\label{2.7}\end{equation}

The gauge invariance is broken by the presence of the boundary in the action $S_{bulk}$ \eqref{2.5}:
\begin{equation}
\delta_{gauge} S_{bulk}=-\alpha\int d^3x\ \delta(x_2)\epsilon_{ij}\lambda\partial_iA_j \ ,
\label{2.8}\end{equation}
where we used $\partial_\mu\theta(x_2)=\delta_{2\mu}\delta(x_2)$. The breaking \eqref{2.8} depends on the CS constant $\alpha$ only since, under gauge transformations, the Chern-Simons Lagrangian transforms  into a total derivative, while the Maxwell Lagrangian is gauge invariant.

The total action we consider is 
\begin{equation}
S_{tot}=S_{bulk}+S_{gf}+S_\gamma+S_{bd} \ ,
\label{2.9}\end{equation}
where $S_{bulk}$ is the bulk MCS action \eqref{2.5} with boundary at $x_2=0$, $S_{gf}$ is the gauge fixing term
\begin{equation}
S_{gf}=\int d^3x\ \theta(x_2)bA_2 \ ,
\label{2.10}\end{equation}
where $b(x)$ is a Lagrange multiplier enforcing the gauge fixing choice $A_2=0$, and $S_\gamma$ is the source term 
\begin{equation}
S_{\gamma}=\int d^3x\ \left[\theta(x_2)\gamma_i A_i +\delta(x_2)\gamma_{2i}(\partial_2A_i)\right]\ ,
\label{2.11}\end{equation}
where $\{\gamma_i, \gamma_{2i}\}$ are external sources coupled to $\{A_i,(\partial_2A_i)\}$, respectively. Notice that an external source is coupled also to the $\partial_2$-derivative of the components $A_i$ of the gauge field, since, on the boundary $x_2=0$, $A_i$ and $(\partial_2A_i)$ must be considered as independent fields.

Finally,  the most general term $S_{bd}$ living on the boundary $x_2=0$ and respecting the power counting assignments \eqref{2.4}, is
\begin{eqnarray}
S_{bd}&=&\int d^3x\ \delta(x_2)\left(
a_1A_0^2
+a_2A_0A_1
+a_3A_1^2
+a_4\partial_0A_0A_1
+a_5\partial_1A_0A_1 \right.
\nonumber \\
&&\left. +a_6(\partial_2A_0)A_0
+a_7(\partial_2A_0)A_1
+a_8(\partial_2A_1)A_0
+a_9(\partial_2A_1)A_1
\right).
\label{2.12}\end{eqnarray}

In the boundary contribution \eqref{2.12} to the total action \eqref{2.9}, terms involving $A_2$ (and its derivatives) have not been introduced, due to the gauge condition $A_2=0$. Moreover, the fact of considering $\left.A_i(x)\right|_{x_2=0}$ and $\left.(\partial_2A_i(x))\right|_{x_2=0}$ as independent fields on the boundary $x_2=0$ does not allow  integration by parts with respect to $x_2$. In \eqref{2.12} the parameters $a_1$, $a_2$ and $a_3$ have mass dimension one, while $a_i$, $i=4,...,9$ are massless. 

We remark that the 2D covariance, in general, is  broken by the boundary term $S_{bd}$ \eqref{2.12}, unless $a_1=a_3$, $a_6=a_9$, $a_7+a_8=0$, and $a_2=a_4=a_5=0$, for which the corresponding Lorentz invariant boundary action reads
\begin{equation}
S_{bd}=\int d^3x\ \delta(x_2)\left(
a_1A_iA_i
+a_6(\partial_2A_i)A_i
+a_7\epsilon_{ij}(\partial_2A_i)A_j
\right).
\label{2.13}\end{equation}
The presence itself of a boundary breaks 3D Lorentz invariance of the action $S_{MCS}$ \eqref{2.1}, and is also violated by the gauge fixing choice $A_2=0$. Nevertheless, 2D Lorentz invariance on the plane $x_2=0$ might still be imposed, but, as in the case of pure CS theory, this turns out to represent a too restrictive constraint for the existence of a nontrivial dynamics on the boundary \cite{Maggiore:2017vjf}, in a similar way to what happens in linearized gravity, where non Lorentz invariant mass terms for the graviton, possibly lying outside the Fierz-Pauli paradigm, have been studied \cite{Rubakov:2004eb,Libanov:2005vu,Blasi:2015lrg,Blasi:2017pkk}. For this reasons, we do not restrict ourselves to a 2D Lorentz invariant boundary term, keeping \eqref{2.12} as the most general one. 


\section{Boundary conditions, Ward identity and continuity equation}

From the total action $S_{tot}$ \eqref{2.9}, the equations of motion read
\begin{eqnarray}
\frac{\delta S_{tot}}{\delta A_0} 
&=& 
\theta(x_2)[2\alpha(\partial_1A_2-\partial_2A_1)
+4\beta(\partial_0\partial_1A_1-\partial_1^2A_0+\partial_0\partial_2A_2-\partial_2^2A_0)
+\gamma_0] 
\nonumber \\
&&
+\delta(x_2)[2a_1A_0+(a_2-\alpha)A_1+4\beta\partial_0A_2+(a_6-4\beta)(\partial_2A_0)
\nonumber\\
&&-a_4\partial_0A_1-a_5\partial_1A_1
+a_8(\partial_2A_1)] 
\label{3.1}\\
\frac{\delta S_{tot}}{\delta A_1} 
&=& 
\theta(x_2)[2\alpha(\partial_2A_0-\partial_0A_2)
+4\beta(-\partial_0^2A_1+\partial_0\partial_1A_0+\partial_1\partial_2A_2-\partial_2^2A_1)
+\gamma_1] 
\nonumber \\
&&
+\delta(x_2)[(a_2+\alpha)A_0+2a_3A_1+4\beta\partial_1A_2+(a_9-4\beta)(\partial_2A_1)
\nonumber\\
&&
+a_4\partial_0A_0+a_5\partial_1A_0 + a_7(\partial_2A_0) ]
\label{3.2}
\end{eqnarray}

The conditions on the boundary $x_2=0$ are obtained by integrating the equations of motion with respect to $x_2$ between zero and an infinitesimal $\epsilon >0$, and then letting $\epsilon\rightarrow 0$. This turns out to be equivalent to putting equal to zero the $\delta(x_2)$-terms in the equations of motion. The result is
\begin{equation}
2a_1A_0+(a_2-\alpha)A_1+(a_6-4\beta)(\partial_2A_0) 
-a_4\partial_0A_1-a_5\partial_1A_1
+a_8(\partial_2A_1) = 0
\label{3.3}\end{equation}
\begin{equation}
(a_2+\alpha)A_0+2a_3A_1+(a_9-4\beta)(\partial_2A_1) 
+a_4\partial_0A_0+a_5\partial_1A_0 + a_7(\partial_2A_0) =0
\label{3.4}\end{equation}

From the equations of motion, the following integrated Ward identity is easily derived
\begin{eqnarray}
\int_0^\infty dx_2\ \partial_i\gamma_i &=&
\left.2\alpha(\partial_1A_0-\partial_0A_1)\right|_{x_2=0}
-4\beta \left[
\partial_0\left.(\partial_2A_0)\right|_{x_2=0} + 
\partial_1\left.(\partial_2A_1)\right|_{x_2=0}
\right] \nonumber\\
&=&
\left.-2\alpha\epsilon_{ij}\partial_iA_j\right|_{x_2=0}
-4\beta\partial_i\left.(\partial_2A_i)\right|_{x_2=0}
\nonumber\\
&=&-2\epsilon_{ij}\partial_i\left[\alpha A_j-2\beta\epsilon_{jk}(\partial_2A_k)\right]_{x_2=0}. 
\label{3.5}\end{eqnarray}

Going on-shell, $i.e.$ putting the external sources $\gamma_i=0$, and denoting with $X=X_i=(x_0,x_1)$ the 2D coordinates on the boundary $x_2=0$, the r.h.s. of \eqref{3.5} yields the 2D conservation  equation
\begin{equation}
\partial_iJ_i(X)=0\ ,
\label{3.6}\end{equation}
where
\begin{equation}
J_i(X)\equiv \epsilon_{ij}\left[\alpha A_j-2\beta\epsilon_{jk}(\partial_2A_k)\right]_{x_2=0}\ .
\label{3.7}\end{equation}
The conservation equation \eqref{3.6} is physically interpreted as the electric continuity equation, and  the two components of the conserved 2D current $J_i(X)$ are the charge density and the current density, respectively: 
\begin{eqnarray}
J_0(X)\equiv\rho(X)&=&\left.\alpha A_1(x)+2\beta(\partial_2A_0)(x)\right|_{x_2=0}\label{3.8}\\
J_1(X)\equiv J(X)&=&-\left.\alpha A_0(x)+2\beta(\partial_2A_1)(x)\right|_{x_2=0}\label{3.9}\ .
\end{eqnarray}
The general solution of the continuity equation \eqref{3.6} with $J_i(X)$ given by \eqref{3.7},  defines a 2D scalar field
\begin{equation}
\left.\alpha A_i(x)-2\beta\epsilon_{ij}(\partial_2A_j)(x)\right|_{x_2=0}\equiv\partial_i\Phi(X)\ .
\label{3.10}\end{equation}
We remark that the relation \eqref{3.10} characterizes the scalar-tensor duality which reveals the presence, on the boundary of fermionic degrees of freedom, allowing a Bose representation for fermionic massless fields \cite{Aratyn:1984jz,Aratyn:1983bg,Amoretti:2013xya}.
The components \eqref{3.8} and \eqref{3.9} of the  2D conserved current $J_i(X)$ \eqref{3.7}, in terms of the 2D scalar field $\Phi(X)$ are
\begin{equation}
\rho(X)=\partial_1\Phi(X)\ \ ;
J(X)=-\partial_0\Phi(X)\ .
\label{3.11}\end{equation}

\section{2D Algebras}

Deriving the Ward identity \eqref{3.5} with respect to $\gamma_k(x')$, putting afterwards $\gamma=0$, which is equivalent to going on-shell, and identifying $x_0$ as the coordinate with respect to which the time ordered product is performed, we have
\begin{eqnarray}
\partial_k\delta^{(2)}(X-X') &=& 
-2\epsilon_{ij}\partial_i \left\langle \left[\alpha A_j - 2\beta\epsilon_{jk} (\partial_2A_k)\right](X)A_k(X')\right\rangle\nonumber\\
&=&-2 \delta(x_0-x_0')\left[ \alpha A_1(X) + 2\beta(\partial_2A_0)(X),A_k(X')\right]\nonumber\\
&&-2\left\langle \epsilon_{ij}\partial_i\left[\alpha A_j - 2\beta\epsilon_{jk} (\partial_2A_k)\right](X)A_k(X')\right\rangle \ .
\label{4.1}\end{eqnarray}
Using the conservation equation  \eqref{3.6}, the last term on the right hand side of \eqref{4.1} vanishes, and we are left with the 2D algebra
\begin{equation}
\partial_1\delta^{(2)}(X-X') =
-2\delta(x_0-x'_0)
\left[
\alpha A_1(X)+2\beta (\partial_2A_0)(X),A_1(X')\right]\ ,
\label{4.2}\end{equation}
where we put $k=1$ in \eqref{4.1}. Analogously, deriving the Ward identity \eqref{3.5} with respect to $\gamma_{2k}(x')$ and going on-shell, we get
\begin{equation}
\delta(x_0-x'_0)\left[\alpha A_1(X)+2\beta(\partial_2A_0)(X),(\partial_2A_k)(X')\right]=0\ .
\label{4.3}\end{equation}
\textcolor{black}{
A linear combination of the relations \eqref{4.2} and \eqref{4.3} with $k=0$, leads to the 2D algebra
\begin{equation}
\partial_1\delta^{(2)}(X-X') =
-2\delta(x_0-x'_0)
\left[
(
\alpha A_1+2\beta (\partial_2A_0))(X),
(
A_1+\frac{2\beta}{\alpha}(\partial_2A_0))(X')\right]\ ,
\label{4.4}\end{equation}
which, in turn, can be written in terms of the 2D scalar field \eqref{3.10} as follows
\begin{equation}
\partial_1\delta^{(2)}(X-X') =
-2\delta(x_0-x'_0)
\left[
\partial_1\Phi(X),
\frac{\partial_1\Phi(X')}{\alpha}
\right]\ .
\label{4.4.1}\end{equation}
The possibility of writing the 2D algebra as \eqref{4.4.1} is crucial, since, as we shall see, in this form the algebra represents the starting point towards the construction of the 2D action for the scalar field induced by the conservation equation \eqref{3.6}. Notice that this is possible only for $\alpha\neq 0$, $i.e.$ in presence of the Chern-Simons term.
}

We stress again that the CS and Maxwell contributions manifest themselves through the dependence on the couplings $\alpha$ and $\beta$, which appear not only in the 2D algebra \eqref{4.4}, but also in the conserved current \eqref{3.7}. For instance, if $\beta=0\ ;\ \alpha \neq 0$, the pure CS case with boundary is recovered, which has been extensively studied \cite{Maggiore:2017vjf,Blasi:1990pf,Emery:1991tf}. The canonical dimension of the gauge field is determined then by the CS term only, and therefore switches to $[A]=1$. As a consequence, the only terms appearing in $S_{bd}$ \eqref{2.12} are the first three, because of power counting. The algebra \eqref{4.4} reduces to a Ka\c{c}-Moody (KM) algebra, with central charge related to the CS coupling $\alpha$.

From the algebra \eqref{4.4}, the cases we are going to study are the following: 
\begin{enumerate}
\item
$(\partial_2A_0)(X)=0$.
\item
$A_1(X)=0$.
\item
$(\partial_2A_0)(X)\neq0$, $A_1(X)\neq0$.
\end{enumerate}

Notice that the case $A_1(X)=(\partial_2A_0)(X)=0$, $i.e.$ case 1. $together\ with$ 2., trivializes the algebra \eqref{4.4},  and therefore is excluded.

\subsection{$(\partial_2A_0)(X)=0$: Neumann boundary condition for  $A_0(x)$ }

Additional constraints in this case are: 
\begin{enumerate}
\item
\begin{equation}
A_1(X)\neq0\ ,
\label{4.5}\end{equation} 
in order to keep the algebra \eqref{4.4} nontrivial.
\item
\begin{equation}
(\partial_2A_1)(X)\neq 0\ ,
\label{4.6}\end{equation} 
in order to maintain a $\beta$-dependence in the conserved current $J_i(X)$ \eqref{3.7}. Notice indeed that imposing Neumann boundary conditions on both fields $A_i(x)$ would eliminate the Maxwell contribution both in the algebra and in the conserved current: it is therefore equivalent to put $\beta=0$ in the bulk action \eqref{2.5}, reducing therefore to the known case of the pure Chern-Simons theory with boundary \cite{Maggiore:2017vjf}.
\end{enumerate}

In this case the KM algebra is recovered
\begin{equation}
\partial_1\delta^{(2)}(X-X')=-2\delta(x_0-x'_0)[\alpha A_1(X),A_1(X')]
\label{4.7}\end{equation}
and the conservation equation \eqref{3.6} is solved introducing a scalar field $\Phi(X)$
\begin{equation}
\alpha A_1(X)=\partial_1\Phi(X)\ \ ;
\alpha A_0(X)-2\beta(\partial_2A_1)(X)=\partial_0\Phi(X).
\label{4.8}\end{equation}
The components \eqref{3.8} and \eqref{3.9} of the conserved current $J_i(X)$ therefore are
\begin{equation}
\rho=\partial_1\Phi\ \ ;\ \  J=-\partial_0\Phi.
\label{4.9}\end{equation}
Notice that, while the algebra \eqref{4.7} is the same KM algebra of the pure CS theory with boundary, with central charge related to the CS coupling, the component $J(X)$ \eqref{3.9} of the 2D conserved current $J_i(X)$ \eqref{3.7}, which is a physical observable, turns out to be nontrivially modified by the Maxwell term.

In terms of the 2D scalar field $\Phi(X)$, the algebra \eqref{4.7} reads
\begin{equation}
\partial_1\delta^{(2)}(X-X')=-2\delta(x_0-x_0')[\partial_1\Phi(X),\frac{\partial_1\Phi(X')}{\alpha}]\  ,
\label{4.10}\end{equation}
from which, factorizing $\partial_1$ on both sides,  we get
\begin{eqnarray}
\delta^{(2)}(X-X') &=& \delta(x_0-x_0') [-2\Phi(X),\frac{\partial_1\Phi(X')}{\alpha}] \label{4.11}\\
&=& \delta(x_0-x_0') [q(X),p(X')]\label{4.12}\ ,
\end{eqnarray}
which leads to the identification of the 2D canonical variables
\begin{equation}
q(X)\equiv-2\Phi(X)\ \ ;\ \ p(X)\equiv \frac{\partial_1\Phi(X)}{\alpha}\ .
\label{4.13}\end{equation}
We are now able to write a 2D action, as the integral of a Lagrangian density given by the sum of $p\dot{q}$ with all possible terms with mass dimension two, without ``time'' derivative $\partial_0$, in order to maintain the identification \eqref{4.13} of the canonical variables $q(X)$ and $p(X)$, together with the translation invariance implicit in the definition of the scalar field $\Phi(X)$ \eqref{4.9}:
\begin{equation}
S_{2D} = \int d^2X\left[\frac{\partial_1\Phi}{\alpha}\partial_0(-2\Phi)+\kappa(\partial_1\Phi)^2\right]
\label{4.14}\end{equation}
where $\kappa$ is a real constant to be determined by asking compatibility with the boundary conditions \eqref{3.3} and \eqref{3.4}, as we shall see shortly. From the 2D action \eqref{4.14} the equation of motion for the scalar field is
\begin{equation}
\frac{\delta S_{2D}}{\delta\Phi}=\frac{4}{\alpha}\partial_1\left[\partial_0\Phi-\frac{\alpha}{2}\kappa\partial_1\Phi\right]=0\ ,
\label{4.15}\end{equation}
which is satisfied by a chiral scalar field propagating on the 2D boundary with velocity 
\begin{equation}
v_{chiral}=- \frac{\alpha}{2}\kappa\ .
\label{4.16}\end{equation}
Let us consider now the compatibility with the boundary conditions \eqref{3.3} and \eqref{3.4}. Remembering that we are considering the case $(\partial_2A_0)(X)=0\ , (\partial_2A_1)(X)\neq0$ and $A_1(X)\neq0$, we must require that the boundary conditions, written in terms of the 2D scalar field $\Phi(X)$ through \eqref{4.8}, are homogeneous in the number of derivatives. This condition is obtained by imposing $a_4=a_5=0$. The boundary conditions reduce finally to 
\begin{eqnarray}
2a_1A_0+(a_2-\alpha)A_1+a_8(\partial_2A_1) &=& 0 \label{4.17}\\
(a_2+\alpha)A_0+2a_3A_1+(a_9-4\beta)(\partial_2A_1) &=& 0 \label{4.18}\ .
\end{eqnarray}
On the other hand, the equation of motion \eqref{4.15} of the scalar field $\Phi(X)$, written in terms of bulk fields $A$, is solved if
\begin{equation}
\frac{4}{\alpha}[\alpha A_0-2\beta(\partial_2A_1)]-2\kappa[\alpha A_1] = 0\ .
\label{4.19}\end{equation}
Compatibility can be reached in many ways, all leading to the same physical result: it is possible to choose the parameters appearing in $S_{bd}$ \eqref{2.12} in order to have the chiral velocity \eqref{4.16} of the scalar field either positive or negative, or null.
For instance, if 
$a_2=\alpha,\ ,\ \kappa=-\frac{2a_3}{\alpha^2}\ ` a_9=0$, the two equations \eqref{4.18} and \eqref{4.19} coincide. 
We are then left with $2a_1A_0+a_8(\partial_2A_1)=0$ which can be satisfied either with  $a_1=a_8=0$, or by imposing a Robin boundary condition $A_0=-\frac{a_8}{2a_1}(\partial_2A_1)\neq0$. 
We have then 
\begin{equation}
v_{chiral}=\frac{a_3}{\alpha}\ ,
\label{4.20}\end{equation}
which, indeed, can be tuned negative, positive or zero. 

\subsection{$A_1(X)=0$: Dirichlet boundary condition for $A_1(X)$ }

The additional constraint to impose  in this case is 
\begin{equation}
(\partial_2A_0)(X)\neq0\ ,
\label{4.21}\end{equation}
in order to have the nontrivial algebra
\begin{equation}
\partial_1\delta^{(2)}(X-X')=-2\left[2\beta(\partial_2A_0)(X),\frac{2\beta}{\alpha}(\partial_2A_0)(X')\right]\ ,
\label{4.22}\end{equation}
which is a KM algebra with central charge $=-\frac{8\beta^2}{\alpha}$, depending, this time on both the CS and the Maxwell couplings.
The components of the conserved current $J_i(x)$ \eqref{3.7} are
\begin{eqnarray}
J_0(X) &=& \partial_1\Phi(X)=2\beta(\partial_2A_0)(X)\label{4.23}\\
J_1(X) &=& -\partial_0\Phi(X)=-\alpha A_0(X)+2\beta(\partial_2A_1)(X)\ ,
\label{4.24}\end{eqnarray}
where \eqref{3.11} has been used to introduce the 2D scalar field $\Phi(X)$.
Written in terms of the scalar field, the algebra \eqref{4.22}  reads
\begin{equation}
\partial_1\delta^{(2)}(X-X')=-\frac{2}{\alpha}\left[\partial_1\Phi(X),\partial_1\Phi(X')\right]\ ,
\label{4.25}\end{equation}
from which we can identify the 2D canonical variables $q(X)$ and $p(X)$
\begin{equation}
\delta^{(2)}(X-X')=\left[-\frac{2}{\alpha}\Phi(X),\partial_1\Phi(X')\right]\equiv\left[q(X),p(X')\right]\ ,
\label{4.26}\end{equation}
and, analogously to what has been done in the previous case, we get the most general 2D action
\begin{equation}
S_{2D} = \int d^2X\left[\partial_1\Phi\partial_0\left(-\frac{2}{\alpha}\Phi\right)+\kappa(\partial_1\Phi)^2\right]\ ,
\label{4.27}\end{equation}
where $\kappa$ is a real constant. The equation of motion  is
\begin{equation}
\frac{\delta S_{2D}}{\delta\Phi}=\frac{4}{\alpha}\partial_1\left[\partial_0\Phi-\frac{\alpha}{2}\kappa\partial_1\Phi\right]=0\ ,
\label{4.28}\end{equation}
which is satisfied by a chiral scalar field with velocity \eqref{4.16}.
The chiral equation, in terms of the bulk fields restricted on the boundary $x_2=0$ $A_i(X)$ is
\begin{equation}
\partial_0\Phi-\frac{\alpha}{2}\kappa\partial_1\Phi=0\Rightarrow
\alpha A_0-2\beta(\partial_2A_1)-\alpha\beta\kappa(\partial_2A_0)=0\ ,
\label{4.29}\end{equation}
which makes easier the study of the compatibility with the boundary conditions \eqref{3.3} and \eqref{3.4} 
\begin{eqnarray}
2a_1A_0+(a_6-4\beta)(\partial_2A_0)+a_8(\partial_2A_1) &=&0\label{4.30}\\
(a_2+\alpha)A_0+a_7(\partial_2A_0)+(a_9-4\beta)(\partial_2A_1) &=&0\label{4.31}\ ,
\end{eqnarray}
where the case at hand $A_1(X)=0$ has been imposed , as well as the homogeneity in the number of derivatives: $a_4=a_5=0$. Two examples of solutions are the following
\begin{eqnarray}
a_1=\frac{\alpha}{2}\ ;\
a_8=-2\beta\ ;\
\kappa=\frac{4\beta-a_6}{\alpha\beta}\ ;\
a_2=-\alpha\ ;\
a_9=4\beta\ ;\
a_7=0\label{4.32}\\
a_2=0\ ;\
a_9=2\beta\ ;\
\kappa=-\frac{a_7}{\alpha\beta}\ ;\
a_1=0\ ;\
a_6=4\beta\ ;\
a_8=0;
\label{4.33}\ .
\end{eqnarray}
As in the previous case, also in this class of solutions $v_{chiral}$ can be greater, less or equal to zero. Notice that, differently from the previous case \eqref{4.20}, for Dirichlet boundary condition for $A_1(X)$, in both solutions \eqref{4.32} and \eqref{4.33} the chiral velocity depends on the Maxwell coupling only. From \eqref{4.16}, we have indeed, respectively for the two above solutions
\begin{equation}
v_{chiral}=\frac{a_6-4\beta}{2\beta} \ ;\ v_{chiral}=\frac{a_7}{2\beta}.
\label{4.34}\end{equation}

\subsection{$(\partial_2A_0)(X)\neq0\ ;\ A_1(X)\neq0$: generic case}

The algebra \eqref{4.4} keeps all its terms, and, written in terms of the scalar fields by means of \eqref{3.10}, reads
\begin{equation}
\partial_1\delta^{(2)}(X-X')  
=
-2\delta(x_0-x_0')\left[\partial_1\Phi(X),\frac{1}{\alpha}\partial_1\Phi(X')\right]\ ,
\label{4.35}\end{equation}
from which
\begin{equation}
\delta^{(2)}(X-X') =\delta(x_0-x_0')\left[-2\Phi(X),\frac{1}{\alpha}\partial_1\Phi(X')\right]\equiv
\delta(x_0-x_0')\left[q(X),p(X')\right]\ ,
\label{4.36}\end{equation}
and
\begin{equation}
p\dot{q}=\frac{1}{\alpha}\partial_1\Phi\partial_0(-2\Phi)\ ,
\label{4.37}\end{equation}
so that, as in the previous cases, the 2D action reads
\begin{eqnarray}
S_{2D} &=&\int d^2X\ (p\dot{q}+\mbox{terms without $\partial_0$}) \nonumber\\
&=&
\int d^2X\ \left[-\frac{2}{\alpha}\partial_0\Phi\partial_1\Phi+\kappa(\partial_1\Phi)^2\right]\ , \label{4.38}
\end{eqnarray}
with $\kappa$ real constant. The equation of motion, satisfied by a 2D chiral scalar $\Phi(X)$, is
\begin{equation}
\frac{\delta S_{2D}}{\delta\Phi}=\frac{4}{\alpha}\partial_1\left[\partial_0\Phi-\frac{\alpha}{2}\kappa\partial_1\Phi\right]=0\ ,
\label{4.39}\end{equation}
and the chiral velocity is again given by \eqref{4.16}.

The ``holographic link'' is obtained by imposing the compatibility between the 2D equation of motion of the 2D scalar field $\Phi(X)$ and the boundary conditions \eqref{3.3} and \eqref{3.4} on the gauge fields $A_i(x)$. In order to do this, we use \eqref{3.10} to write the chiral condition induced by the equation of motion \eqref{4.39} as
\begin{equation}
\frac{4}{\alpha}(\alpha A_0-2\beta(\partial_2A_1))-2\kappa(\alpha A_1+2\beta(\partial_2A_0))=0\ .
\label{4.40}\end{equation}
For instance, comparing \eqref{4.40} with the sum of the boundary conditions \eqref{3.3} and \eqref{3.4}, we get  
\begin{eqnarray}
2\alpha &=& 2a_1+a_2+\alpha \\
-4\beta &=&a_8+a_9-4\beta \\
-\alpha^2\kappa&=&a_2-\alpha+2a_3 \\
-2\alpha\beta\kappa&=&a_6-4\beta+a_7\ .
\end{eqnarray}
Again, In the boundary conditions \eqref{3.3} and \eqref{3.4}, the terms $(a_4,a_5)$ are decoupled, and therefore must vanish alone: $a_4=a_5=0$. One of the possible solutions is 
\begin{equation}
a_8+a_9=a_6+a_7=0\ ;\
a_2=\alpha-2a_1\ ;\
a_3=a_1-\alpha\ ;\
\kappa=\frac{2}{\alpha}\ ,
\label{4.45}\end{equation}
which guarantees compatibility between the equation of motion of the 2D scalar field $\Phi(X)$ \eqref{4.39} and the boundary conditions \eqref{3.3} and \eqref{3.4} of the 3D MCS theory with boundary.
According to this particular solution, the boundary conditions become
\begin{equation}
2a_1A_0-2a_1A_1+(a_6-4\beta)(\partial_2A_0)+a_8(\partial_2A_1) = 0 \label{4.46}
\end{equation}
\begin{equation}
2(\alpha-a_1)A_0+2(a_1-\alpha)A_1-a_6(\partial_2A_0)-(a_8+4\beta)(\partial_2A_1) =0\ , \label{4.47} 
\end{equation}
whose sum is
\begin{equation}
2\alpha A_0-2\alpha A_1-4\beta(\partial_2A_0)-4\beta(\partial_2A_1)=0\ ,
\label{4.48}\end{equation}
that is 
\begin{equation}
\alpha A_0 -2\beta(\partial_2A_1)  =  \alpha A_1+2\beta(\partial_2A_0)=0\ ,
\label{4.49}\end{equation}
so that the algebra \eqref{4.4} can equivalently be written as
\begin{equation}
\partial_1\delta^{(2)}(X-X') =
-2\delta(x_0-x'_0)
\left[
(
\alpha A_0-2\beta (\partial_2A_1))(X),
(
A_0-\frac{2\beta}{\alpha}(\partial_2A_1))(X')\right].
\label{4.50}\end{equation}
Notice that these equations can be written in a covariant way
\begin{equation}
\alpha A_i-2\beta\epsilon_{ij}(\partial_2A_j)=0\ ,
\label{4.51}\end{equation}
which therefore represents the boundary conditions on the fields in the generic case treated in this subsection. Notice that \eqref{4.51} can be seen as a kind of twisted Robin boundary condition.
In terms of scalar fields, \eqref{4.51} is
\begin{equation}
\partial_0\Phi=\partial_1\Phi\ .
\label{4.52}\end{equation}
The equation of motion of $\Phi$ is
\begin{equation}
\partial_1(\partial_0\Phi-\partial_1\Phi)=0\ .
\label{4.53}\end{equation}
For this solution, the velocity of the chiral field is fixed: 
\begin{equation}
v_{chiral}=-1\ ,
\label{4.54}\end{equation}
to be compared with the cases studied in the previous subsections 4.1 and 4.3, where the chiral velocity can assume any value: \eqref{4.20} and \eqref{4.34} respectively.

\section{Conclusions}
 In this paper the MCS theory with a single-edged planar boundary has been considered. From the field theoretical point of view, the problem is of the bulk/boundary type, rather than of the left/right hand side (with respect to a boundary) type. Hence, Symanzik's approach based on the separability constraint on the propagators is not well suited, and the situation resembles the holography duality between bulk and boundary. The starting point to realize the lower dimensional dual theory is the bulk Ward identity, broken by the presence of the boundary. The breaking, once the external sources are put equal to zero (which is equivalent to going on-shell), describes a conserved current, which can be identified with the continuity equation for the transport of electric charge. The broken Ward identity represents the holographic link between the 3D bulk theory and its 2D boundary counterpart. In fact, the breaking/continuity equation can be solved by means of a 2D scalar field, which identifies the boundary degree of freedom. On the other hand, the algebra derived from the broken Ward identity, written in terms of the 2D scalar field, can be read as the commutation relations between canonically conjugate variables. The boundary 2D action, dual to the 3D bulk theory, is the most general one which reproduces the canonical commutation relations. This request does not uniquely fix the action. The holographic link is completed by the additional request that the equation of motion for the 2D scalar field is compatible with the boundary conditions for the 3D gauge fields, once these are written in terms of the scalar field determined by the continuity equation. The physical content of this constructions resides in the edge excitations, which are chiral currents whose chiral velocity depends on the parameters of the bulk theory. This agrees with the results obtained for CS theory with boundary, where the chiral velocity is proportional to the CS coupling constant. What is peculiar of the case studied in this paper, is that, in presence of a single-edged boundary, the chiral velocity depends on both the coupling constants of the bulk theory (the CS and the Maxwell ones). In particular, dependence is found also on the non-topological Maxwell coupling, which is transparent in the case of doubled-edged boundary, treated {\it \`a la} Symanzik \cite{Blasi:2010gw}. In addition, and remarkably, the dependence of the chiral velocity on the two coupling constants of the bulk MCS theory is tightly related to the possible boundary conditions on the 3D gauge fields according to the following simple rule: for Neumann boundary conditions the chiral velocity depends on the CS coupling only, for Dirichlet boundary conditions, on the contrary, the chiral velocity depends on the Maxwell coupling only. A third possibility occurs: for twisted Robin boundary conditions, the compatibility equations can be written in a 2D covariant way, and the chiral velocity is uniquely fixed, differently from the previous cases, where the chiral velocity can be tuned to be positive, negative, or even null.

\vspace{1cm}

{\bf Acknowledgements}

The support of INFN Scientific Initiative SFT: ``Statistical Field Theory, Low-Dimensional Systems, Integrable Models and Applications'' is acknowledged.


\end{document}